\documentstyle[12pt]{article}
\begin{document}
\setlength{\unitlength}{1mm}
{\hfill   August 1994 } \vspace*{2cm} \\
\begin{center}
{\Large\bf On "Non-Geometric" Contribution To The Entropy Of
 Black Hole Due To Quantum Corrections.}
\end{center}
\begin{center}
{\large\bf  Sergey N.~Solodukhin$^{\ast}$}
\end{center}
\begin{center}
{\bf Bogoliubov Laboratory of Theoretical Physics, Joint Institute for
Nuclear Research, Head Post Office, P.O.Box 79, Moscow, Russia}
\end{center}
\vspace*{2cm}
\begin{abstract}
The quantum corrections to the entropy of charged black holes are calculated.
The Reissner-Nordstrem and dilaton black holes are considered. The appearance
of
logarithmically divergent  terms not proportional to the horizon area
is demonstrated. It is shown that the complete entropy which is sum
of classical Bekenstein-Hawking entropy and the quantum correction is
proportional
to the area of quantum-corrected horizon.
\end{abstract}
\begin{center}
{\it PACS number(s): 04.60.+n, 12.25.+e, 97.60.Lf, 11.10.Gh}
\end{center}
\vskip 1cm
\noindent $^{ \ast}$ e-mail: solod@thsun1.jinr.dubna.su
\newpage
\baselineskip=.8cm

The classical Bekenstein-Hawking entropy of four-dimensional black hole is
known to be proportional to the area of horizon:
\begin{equation}
S_{BH}={ 1 \over 4}{A_h \over \kappa},
\end{equation}
where $\kappa$ is gravitational constant [1]. Roughly speaking, the horizon
is two-dimensional surface which separates the whole space on two different
regions
the free exchange of information between which is impossible. Thus,
the outside observer does not have information about states of quantum field in
the region
inside the horizon and therefore must trace over all such states. The entropy,
characterizing this  unknowledge, turns out to be determined only by geometry
of the surface separating these two regions, namely it is proportional to
the area of the surface. This fact occurs to be not feature of only
gravitational
objects but rather typical [2].

It is reasonable to ask whether this geometric character of the black hole
entropy
remains valid when the quantum corrections (say, due to quantum
fluctuations of matter fields in the black hole background) are taken into
account.

Approximating the metric of black hole of infinitely large mass by
more simple Rindler metric, it was shown [3] that quantum correction to (1)
again takes the geometric character:
\begin{equation}
S^q={1 \over 48\pi} {A_h \over \epsilon^2}
\end{equation}
though it is divergent when the ultraviolet cut-off $\epsilon$ tends to zero.
This divergence was related with information loss in the black hole [4].

However, recently  [5] we shown that the Rindler metric is not good model for
the black hole space-time.
 The reason is that the horizon surfaces of black hole (sphere) and
 Rindler space (plane) are topologically different.
 For black hole of finite mass at the same time with (2)
one also observes the logarithmically divergent mass independent term:
\begin{equation}
S^q={1 \over 48\pi} {A_h \over \epsilon^2}+{1 \over 45} \log {\Lambda \over
\epsilon},
\end{equation}
where $\Lambda$ is infra-red cut-off.
This term does not have the geometric character and resembles the quantum
correction
to entropy of two-dimensional black hole [6]. Generally, entropy is defined up
to
arbitrary additive constant. Hence, one could assume that this term
is not essential and does not influence the physics. In this note we show
(on example of charged black holes) that
appearance of such, non-geometric, logarithmically divergent terms
is typical in four dimensions. In general case, these terms depend on the
characteristics of the black hole (charge, mass, etc.) and therefore
can not be neglected as non-essential additive constants. We use the path
integral method of Gibbons and Hawking [7] to calculate the corrections to
entropy of black hole. The basic formulas can be found in [5].

\bigskip

In the Euclidean path integral approach to statistical field system taken under
temperature $T=(2\pi\beta)^{-1}$ one considers the fields which are
periodical with respect to imaginary time $\tau$ with period $2\pi\beta$.
For arbitrary $\beta$ the classical black hole metric is known to have
conical singularity which disappears only for special Hawking inverse
temperature $\beta_H$.

Let matter is described by the action:
\begin{equation}
I_{mat}={1 \over 2} \int_{}^{}(\nabla \Phi)^2\sqrt{g}d^4x
\end{equation}
Then contribution to the energy and entropy due to matter fluctuations is
given by
\begin{equation}
E^q={1 \over 2\pi}\partial_\beta I_{eff}(\beta,\Delta)|_{\beta=\beta_H}
\ \ \ S^q=(\beta \partial_\beta -1) I_{eff}(\beta,\Delta)|_{\beta=\beta_H},
\end{equation}
where $\Delta=\nabla_\mu \nabla^\mu$ is the Laplace operator; $I_{eff}(\beta,
\Delta)={1 \over 2} \ln det \Delta_{g_\beta}$ is the one-loop
effective action calculated in the classical black hole background with conical
singularity at the horizon. In order to take derivative $\partial_\beta$
in (5) we assume that $\beta$ is slightly different of $\beta_H$.

The logarithm of determinant in the De Witt-Schwinger proper time
representation is as follows:
\begin{equation}
\log det \Delta =-\int_{\epsilon^2}^{\infty}s^{-1} Tr (e^{-s\Delta}),
\end{equation}
where the integral over $s$ is cutted on the lower limit under
$\epsilon^2=L^{-2}$,
$L$ is maximal impulse.

In four dimensions we have the asymptotic expansion:
\begin{equation}
Tr (e^{-s \Delta})={1 \over (4\pi s)^2} \sum_{n=0}^{\infty} a_n s^n
\end{equation}
The divergent part of the effective action is given by
\begin{equation}
I_{eff}=-{1 \over 32 \pi^2}({1 \over 2} a_0 \epsilon^{-4} +a_1
 \epsilon^{-2} +a_2 \log ({\Lambda \over \epsilon})^2)
\end{equation}

\bigskip

The black hole metric in vicinity of horizon ($\rho=0$) has the form:
\begin{equation}
ds^2=\alpha^2 (\rho^2+ C \rho^4)d\phi^2+d\rho^2+(\gamma_{ij}(\theta)+
h_{ij}(\theta)\rho^2)d\theta^id\theta^j
\end{equation}
where $C=const$, $\alpha={\beta \over \beta_H}$ and we introduced new
coordinate
$\phi =\beta^{-1}\tau$ which has period $2\pi$. Near the horizon ($\rho=0$)
this Euclidian space looks as direct product $M_\alpha=C_\alpha \otimes
\Sigma$.
$C_\alpha$ is two dimensional cone with metric $ds^2=\alpha^2\rho^2d\phi^2+
d\rho^2$, $\Sigma$ is the horizon surface with metric $\gamma_{ij}(\theta)$.
It was shown recently [8]  that for background
like this the coefficients in the expression (7) take the form
\begin{equation}
a_n=a^{reg}_n +a_{\alpha,n}
\end{equation}
where $a_n^{reg}$ are standard  coefficients $a_n=\int_{M_\alpha}^{}a_n(x,x)
d\Omega (x)$, given by the integrals over the smooth domain of $M_\alpha$;
the coefficients $a_{\alpha,n}$ are  surface terms determined by integrals
over the horizon $\Sigma$:
\footnotemark\
\addtocounter{footnote}{0}\footnotetext{Our convention for the curvature and
Ricci tensor is $R^\alpha_{\ \beta\mu\nu}=\partial_\mu \Gamma^\alpha_{\
\nu\beta}-...,$
and $R_{\mu\nu}=R^\alpha_{\ \mu\alpha\nu}$.}
\begin{eqnarray}
&&a_{\alpha,0}=0; \ \ \ a_{\alpha , 1}={\pi \over 3}{(1-\alpha)(1+\alpha) \over
\alpha}
\int_{\Sigma}^{}\sqrt{\gamma} d^2 \theta \ ; \nonumber \\
&&a_{\alpha , 2}={\pi \over 18} {(1-\alpha)(1+\alpha) \over \alpha}
\int_{\Sigma}^{}R \sqrt{\gamma} d^2 \theta \nonumber \\
&&-{\pi \over 180}
{(1-\alpha)(1+\alpha)(1+\alpha^2) \over \alpha^3}
\int_{\Sigma}^{}(R_{\mu\nu}n^\mu_i n^\nu_i -2R_{\mu\nu\alpha\beta}n^\mu_i
n^\alpha_i
n^\nu_j n^\beta_j )\sqrt{\gamma} d^2 \theta
\end{eqnarray}
where $n^i$ are two vectors orthogonal to surface $\Sigma$ ($n^\mu_i n_{j}^\nu
g_{\mu\nu}=
\delta_{ij}$). For metric (9) we may take $n^\mu_1=((\alpha \rho)^{-1},
0,0,0)$, $n^\mu_2=(0,1,0,0)$.

For metric (9) we obtain at $\rho=0$:
\begin{eqnarray}
&&R=R_\Sigma -6C-4\gamma^{ij}h_{ij} \nonumber \\
&&R_{\mu\nu}n^\mu_i n^\nu_i-2R_{\mu\nu\alpha\beta}n^\mu_i n^\alpha_i
n^\nu_j n^\beta_j =6C-2\gamma^{ij}h_{ij},
\end{eqnarray}
where $R_\Sigma$ is scalar curvature determined with respect to two-dimensional
metric
$\gamma_{ij}$.

Inserting (10), (11), (12) into (5) we obtain for correction to the entropy
\begin{equation}
S^q={A_\Sigma \over 48\pi \epsilon^2} +({1 \over 18} -{ 1 \over 16\pi}
\int_{\Sigma}^{}({4 \over 5}C+
\frac{2}{5}\gamma^{ij}h_{ij})\sqrt{\gamma}d^2\theta)
\log {\Lambda \over \epsilon}
\end{equation}
where we used the fact that the horizon surface $\Sigma$ is sphere and hence
$\frac{1}{4\pi}\int_{\Sigma}^{}R_{\Sigma}\sqrt{\gamma}d^2\theta=2$;
$A_\Sigma=\int_{\Sigma}^{}\sqrt{\gamma}d^2\theta$ is the horizon area.

If we start with black hole metric written in the Schwarzschild like form
\begin{equation}
ds^2=\beta^2g(r)d\phi^2+{1 \over g(r)} dr^2+ r^2
\tilde{g}_{ij}(\theta)d\theta^i
d\theta^j
\end{equation}
where $\tilde{g}_{ij}(\theta)$ is metric of 2D sphere,
then introducing new radial coordinate
$\rho=\int_{}^{}g^{-1/2}dr$ we obtain  in vicinity of horizon
(which is determined as simple zero of $g(r)$) the metric in the form (9) where
$\gamma_{ij}=r^2_h \tilde{g}_{ij}$, $h_{ij}={r_h \over \beta_H} \tilde{g}_{ij}$
and $C=\frac{1}{6}g''_r|_{r_h}$; $r_h$ is radius of the horizon sphere.
The corresponding Hawking temperature is $\beta_H=2(g'_r(r_h))^{-1}$.

Finally we get for the quantum correction to the entropy:
\begin{equation}
S^q={A_\Sigma \over 48\pi \epsilon^2} +(\frac{1}{18}-{A_\Sigma \over
20\pi}(\frac{1}{6}
g''_r|_{r_h}+{1 \over r_h\beta_H})) \log {\Lambda \over \epsilon}
\end{equation}

We see that logarithmic term in (15) is formally proportional to the horizon
area
$A_\Sigma$. However, the coefficient of proportionality depends
on the background black hole geometry and therefore the whole expression
does not take the form (1).

Let us consider some particular examples.

\bigskip

{\it Example 1. Reissner-Nordstrem black hole.}

The charged black hole is described by metric (14) with
\begin{equation}
g(r)=1 -\frac{2M}{r}+\frac{Q^2}{r^2}, \ \ M \geq Q.
\end{equation}
The largest horizon is located at
\begin{equation}
r_h=M+\sqrt{M^2-Q^2}
\end{equation}
The corresponding Hawking inverse temperature is
\begin{equation}
\beta_H={r^2_h \over \sqrt{M^2-Q^2}}
\end{equation}
and for the second derivative $g''$ we have
\begin{equation}
g''|_{r_h}=2r^{-4}_h(3Q^2-2M^2-2M \sqrt{M^2-Q^2})
\end{equation}
Inserting (17)-(19) into (15) we obtain for the quantum correction to the
entropy:
\begin{equation}
S^q={A_\Sigma \over 48\pi \epsilon^2}+({1 \over 18}-{M \over 15r_h}) \log
{\Lambda \over \epsilon}
\end{equation}
When mass $M$ becomes infinitely large, $r_h \rightarrow 2M$ and (20) coincides
with (3). It is interesting to note that for extreme black hole
($M=Q$, $\beta_H=\infty$, $r_h=M$)  the second term in (20) becomes negative:
\begin{equation}
S^q={A_\Sigma \over 48\pi \epsilon^2}-{1 \over 90} \log {\Lambda \over
\epsilon}
\end{equation}
However, this does not mean that the whole expression (21) is negative since in
limit
$\epsilon \rightarrow 0$ the first positive term is dominant.

\bigskip

{\it Example 2. Dilaton charged black hole}.

The metric of dilaton black hole having electric charge $Q$ and magnetic charge
$P$ takes the form [9]:
\begin{equation}
ds^2=gdt^2+g^{-1}dr^2+R^2d\Omega
\end{equation}
with metric function
\begin{equation}
g(r)={(r-r_+)(r-r_-) \over R^2}, \ \ \ R^2=r^2-D^2
\end{equation}
where $D$ is the dilaton charge: $D={P^2-Q^2 \over 2M}$. The outer and the
inner
horizons are defined as follows
\begin{equation}
r_{\pm}=M\pm r_0 \ ; \ \ r^2_0=M^2+D^2-P^2-Q^2
\end{equation}
Near the outer horizon we have

$$
R^2=R^2_++\frac{r_+}{\beta_h}\rho^2, \ \ R^2_+=r^2_+-D^2
$$

and the metric (22) takes the form (9). The Hawking temperature $\beta_H$ is
\begin{equation}
\beta_H={2(r^2_+-D^2) \over (r_+ -r_-)}
\end{equation}
and for the second derivative $g''_r$ we have
\begin{equation}
g''_r(r_h)={2 \over (r^2_+-D^2)^2} (r^2_+-D^2-2(r_+-r_-)r_+)
\end{equation}
{}From general expression (15) we get for this type of black hole
\begin{equation}
S^q={A_\Sigma \over 48\pi \epsilon^2}+(-{1 \over 90}
 +{2 \over 15}{r_+(r_+-r_-) \over (r^2_+-D^2)}+{1 \over 10}{(r_+-r_-) \over
r_+})
 \log {\Lambda \over \epsilon}
\end{equation}
where $A_\Sigma=4\pi(r^2_+-D^2)$.

It is instructive to consider the black hole with only electric charge
($P=0$). Then $r_0=M-{Q^2 \over 2M}$ ($2M^2>Q^2$), ${r_+(r_+-r_-) \over
(r^2_+-D^2)}=1-{Q^2 \over 4M^2}$ and expression (27) takes the form
\begin{equation}
S^q={A_\Sigma \over 48\pi \epsilon^2}+({1 \over 18}
 +{1 \over 15}({(2M^2-Q^2) \over 2M^2}-{1 \over 5}({2M^2-Q^2 \over 4M^2-Q^2}))
 \log {\Lambda \over \epsilon}
\end{equation}
For large $M$ we again obtain result (3).

In the case of dilaton extreme black hole, $2M^2=Q^2$, the horizon area
vanishes,
$A_\Sigma=0$, and the whole black hole entropy is determined only by the
logarithmically divergent term
\begin{equation}
S^q_{extr}={1 \over 18} \log {\Lambda \over \epsilon}
\end{equation}
Notice that (29) is positive. Expression (29) is very similar to the
entropy of two-dimensional black hole [6]. This can be considered as an
additional
justifying the point
that the dilaton extreme black hole is effectively two-dimensional that was
widely
exploted recently [10].

\bigskip

Thus, we demonstrated on number of examples the appearance of logarithmically
divergent terms in quantum correction to the entropy which are not proportional
to
the horizon area. One could conclude from this that the classical law (1)
is broken due to the quantum corrections. However, one can show that the
complete
black hole entropy
\begin{equation}
S=S_{BH}+S^q
\end{equation}
which is sum of classical Bekenstein-Hawking entropy (1) and the quantum
correction again takes the form (1) being defined with respect to the
renormalized
quantities. The renormalized gravitational constant $\kappa_{ren}$ is
determined as follows [5]:
\begin{equation}
\frac{1}{\kappa_{ren}}=\frac{1}{\kappa}+\frac{1}{12\pi \epsilon^2}
\end{equation}
Then (30) can be written in the form similar to (1):
\begin{equation}
S=\frac{1}{4\kappa_{ren}} A_{\Sigma , \ ren}
\end{equation}
if we define the  quantum corrected radius of horizon  $r_{h, \ ren}$ as
follows
\begin{equation}
4\pi r^2_{h, \ ren}=4\pi r^2_h +\eta l^2_{pl}
\end{equation}
where $l^2_{pl}=\kappa_{ren}$ is the Planck length; quantity $\eta=\eta (M,Q)
\log {\Lambda \over \epsilon}$ absorbs the logarithmic divergence of (30)
and in general depends on bare black hole characteristics: $M, \ Q$, etc.
For the Schwarzschild black hole $\eta$ is positive constant. The expression
like
(33) appears in the work of York [11] describing the quantum  fluctuations of
the
horizon and recently in [12] as result of
quantum deformation of the Schwarzschild solution.
On the other hand, for the charged Reissner-Nordstrem black hole we have
$\eta=(\frac{2}{9}-\frac{4M}{15r_h})\log \frac{\Lambda}{\epsilon}$ and for
the extreme black hole ($Q=M$) $\eta$ is negative.

Expression (33) means that quantum corrections result in the shifting the
horizon
radius by the Planck distance. For the charged black hole with
$Q<M<\sqrt{\frac{25}{24}}Q$
the quantum corrections decrease the horizon radius while for
$M>\sqrt{\frac{25}{24}}Q$
it is increasing. The quantum corrected entropy is determined then with
respect to this quantum corrected horizon in such a way that the law
(1) remains valid. For massive black hole ($M>>M_{pl}$) this shifting of
horizon is negligible. However, it becomes essential and important for
black hole of the Planck mass.

One of the reasons for (33) to be hold could  be the renormalization of mass of
the black hole
that can be calculated in principle from (5). One must take into
account the boundary terms in the effective action which contribute to
the energy (in (8) we neglected such a boundary terms). On the other hand,
(33) can be considered as a result of deformation on small distances of the
Schwarzschild solution due to quantum corrections (see [12]).

Concluding, we calculated the quantum correction to the entropy of charged
black holes
and demonstrated the appearance of logarithmically divergent terms not
proportional
to the horizon area. Nevertheless, we shown that the complete entropy which
is sum of classical Bekenstein-Hawking entropy and the quantum correction
is proportional to the area of the renormalized horizon.

This work is supported in part by the grant RFL000 of the International Science
Foundation.

\end{document}